\documentstyle[aps,multicol,epsf]{revtex}

\input{epsf}
\epsfverbosetrue
\epsfclipon

\newcommand{\putfigxsz}[4]{
   \begin{center}\mbox{\epsfxsize #4
   \epsffile{#1}}
   \end{center}
   }

\begin{document}
\draft
\title{Re-entrant insulator--metal--insulator transition at {\it B}=0
 in a two dimensional hole gas}
\author{A.R. Hamilton, M.Y. Simmons, M. Pepper, E.H. Linfield,
 P.D. Rose, and D.A. Ritchie}
\address{Cavendish Laboratory, Madingley Road, Cambridge CB3 OHE,
 U.K.}
\date{August 1, 1998}
\maketitle
\begin{abstract}
We report the observation of a re-entrant insulator--metal--insulator
transition at $B$=0 in a two dimensional (2D) hole gas in GaAs at
temperatures down to 30~mK. At the lowest carrier densities the holes
are strongly localised. As the carrier density is increased a metallic
phase forms, with a clear transition at $\sigma
\simeq 5e^2/h$. Further increasing the density weakens the metallic
behaviour, and eventually leads to the formation of a second
insulating state for $\sigma \gtrsim 50e^2/h$. In the limit of high
carrier densities, where $k_{F}l$ is large and $r_s$ is small, we thus
recover the results of previous work on weakly interacting systems
showing the absence of a metallic state in 2D.
\end{abstract}


\begin{multicols}{2}
\narrowtext

Evidence has recently emerged pointing towards the existence of a
metallic state in 2D systems with low disorder, in apparent
disagreement with earlier experiments~\cite{Uren} and theoretical
expectations~\cite{Gang4-Gorkov,AA}. In these earlier works it was
shown that logarithmic corrections to the metallic conductivity
existed at low temperatures in low mobility systems, and it was
generally believed that these corrections would also be found at
lower, unobtainable, temperatures in higher mobility samples. Recent
experiments, using electrons in silicon inversion layers~\cite{Krav}
and holes in both SiGe~\cite{Coleridge} and
GaAs~\cite{SimmonsT129,Hanein98,Simmons98} heterostuctures, have shown
a transition from strong localisation at conductances $\sigma < e^2/h$
to metallic behaviour with increasing carrier concentration,
characterised by a significant drop in resistance at low temperatures.

The metallic state has only been observed in 2D systems at low carrier
densities, where electron-electron interactions are known to be
strong, although it is not known if it persists to the very lowest
temperatures. If the metallic state is indeed due to the strength of
the interactions then it might be expected that as the carrier density
is further increased these interactions will no longer dominate and
the metallic phase will cease to exist. Whilst the first transition
from the 2D metallic phase to an insulating state with
\emph{decreasing} carrier density has been widely
observed~\cite{Krav,Coleridge,SimmonsT129,Hanein98,Simmons98,Popovic97,Pudalov97},
there has been no equivalent observation of a second transition from
metallic to insulating behaviour with \emph{increasing} carrier
density, which must occur when $k_{F}l$ is large and
$r_s\rightarrow0$~\cite{Gang4-Gorkov,AA}. Here $k_F$ is the Fermi
wavevector, $l$ is the elastic mean free path, and $r_s$ is the ratio
of the average interparticle Coulomb energy to the Fermi energy.

In this paper we present experimental evidence for both
insulator--metal--insulator transitions as the carrier density is
increased in a two dimensional GaAs hole gas. The observation of a
second metal--insulator transition at $\sigma
\gtrsim 50e^2/h$ ($k_F{l}
\gtrsim 50, r_s \lesssim 8)$ thus bridges the gap
between current experimental data from strongly interacting systems
and earlier studies which showed the absence of a metallic state in a
weakly interacting 2D system~\cite{Uren,Gang4-Gorkov,AA}. The finding
of insulating behaviour at high densities has significant implications
concerning the ultimate low temperature nature of the metallic phase.
Finally we show that the application of a parallel magnetic field
weakens the metallic state, and we map out the $p_s-B_{\parallel}$
phase space over which the 2D metal exists.

\begin{figure}
\putfigxsz{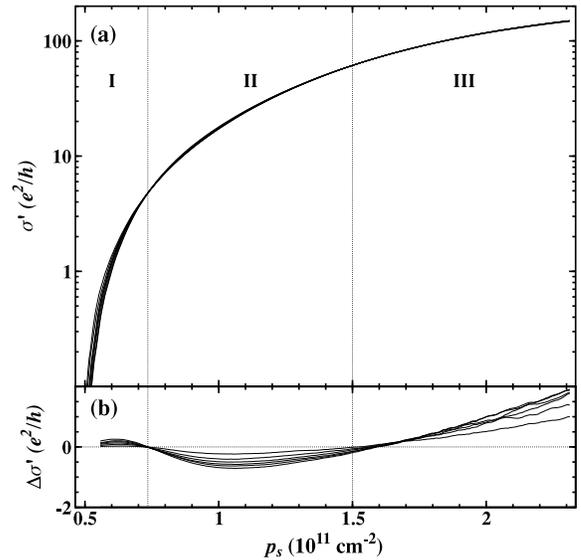}{lab}{cap}{7.5cm}
\caption{(a) Differential conductivity, $\sigma' \propto |\partial{I}/\partial{V}|_E$
for electric fields of 0, 1.5, 3, 4.5, 6, 7.5, and 10.5 mV/cm, as a
function of carrier density ($T=30$~mK). (b) Change in the
differential conductivity $\Delta\sigma'(E)=\sigma'(E)-\sigma'(E{=}0)$
as a function of carrier density. }
\label{fig1}
\end{figure}

The heterostructure used in this study is similar to that described in
Ref.~\cite{Simmons98}, consisting of a 200~{\AA} GaAs quantum well
with an \emph{in-situ} back gate buried 3300~{\AA} below. After
illumination with a red LED the hole density could be varied from
$0-2.3{\times}10^{11}~\text{~cm}^{\text{-2}}$ ($r_s>5$). The peak
mobility, obtained at the highest density, was $1.6{\times}10^{5}
~\text{~cm}^{\text{2}}\text{V}^{\text{-1}}\text{s}^{\text{-1}}$, which
is 30\% lower than the mobility at the same $p_s$ in
Ref.~\cite{Simmons98}. Hall bars of size 450 by 50~$\mu$m were defined
along the $[\bar2 33]$ direction, and data was taken using a dilution
fridge with a base temperature of $\approx$30~mK. Standard four
terminal low frequency (4~Hz) ac lockin techniques were used, with
excitations in the range 33-50~$\mu$V and 0.05-5~nA for constant
voltage and constant current measurements respectively.

Figure~\ref{fig1}(a) demonstrates the presence of the
insulator-metal-insulator transition by plotting the differential
conductivity of the hole gas ($\sigma'(E) \propto
|\partial{I}/\partial{V}|_E$) as a function of carrier density for
different d.c. electric fields $E$ applied along the channel. The
electric field raises the temperature of the holes above the lattice
temperature, making it possible to distinguish between insulating and
metallic states. Insulating behaviour in regions I and III is
characterised by an increase in the conductivity (or differential
conductivity) with increasing temperature (or electric field), whereas
a decrease in $\sigma$ with $E$ or $T$ in region II indicates metallic
behaviour. Whilst the I$\rightarrow$II transition at low carrier
densities can be clearly seen, the second II$\rightarrow$III
transition is less visible since the change in $\sigma'$ due to the
electric field is small compared to the value of $\sigma'$ at the
second transition. The two transitions can be more clearly resolved by
examining the change in the differential conductivity $\Delta\sigma'$
due to the electric field (Fig.~\ref{fig1}(b)). We can thus divide the
graph into three regimes, indicated by the dotted lines:
\begin{enumerate}
  \item[I.] For densities below
  $7.3{\times}10^{10}\text{~cm}^{\text{-2}}$ the sample is in the
  regime of strong localisation: $\sigma$ increases with increasing
  $|E|$, and $\Delta\sigma' > 0$.

  \vspace{-1 ex} \item[II.] An electric field (\emph{i.e.} hole
  temperature) independent point at $\sigma' \simeq 5e^2/h$ ($p_s =
  7.3{\times}10^{10}\text{~cm}^{\text{-2}}, r_s \simeq 9$
  \cite{rsnote}) separates the strongly insulating regime from the
  metallic regime, in which $\Delta\sigma' < 0$.

  \vspace{-1 ex} \item[III.] For densities
  $\gtrsim1.5{\times}10^{11}\text{~cm}^{\text{-2}}$ ($r_s \lesssim 6$)
  the sample once again becomes insulating with $\Delta\sigma' > 0$,
  although the transition between the metallic phase and this second
  insulating phase at $\sigma \approx 50e^2/h$, is not as well defined
  as the first transition. Closer inspection shows that the transition
  moves to higher densities with increasing $E$, the significance of
  which will be discussed later.
\end{enumerate}

This re-entrant insulator--metal--insulator behaviour was also
observed in the temperature dependence of $\sigma$, with a clear $T$
independent transition at $\sigma \simeq 5e^2/h$
($50~\text{mK}<T<1~\text{K}$), and a second, broader, transition at
$\sigma \approx 50e^2/h$.

\begin{figure}
\putfigxsz{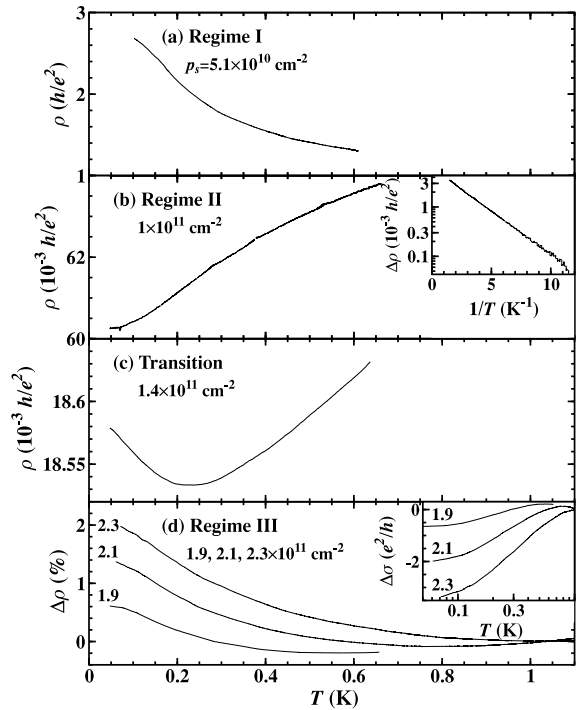}{lab}{cap}{7.5cm}
\caption{Temperature dependence of the resistivity
in the different regimes. (a) Strongly localised regime: $\rho$ shows
hopping conduction. (b) Metallic regime: $\rho$ decreases
exponentially as $T$ is reduced. The inset shows an Arrhenius plot of
$\Delta\rho=\rho(T)-\rho_0$. (c) Transition between regimes II and
III: The metallic behaviour is weaker at this higher density, with
insulating behaviour visible at $T<0.2$~K. (d) Regime III: Fractional
change in resistivity for 3 different carrier densities. The inset
shows the equivalent change in conductivity on a semi-log axis. }
\label{Tdeps}
\end{figure}

In order to determine the nature of this new insulating phase at high
carrier densities we now examine the three phases in more detail, and
plot the temperature dependence of the resistivity for each regime in
Figure~\ref{Tdeps}. At the lowest carrier densities, where
$\sigma<e^2/h$, the system is strongly insulating and there is a large
increase in resistivity with decreasing $T$ (Fig. ~\ref{Tdeps}(a)). In
this regime, close to the transition, $\rho$ follows the form for Mott
variable range hopping: $\rho(T)=\rho_M
\exp[(T/T_0)^{-1/3}]$.
This behaviour is consistent with that observed in previous work on a
slightly higher quality sample, where $T^{-1/2}$ behaviour was
observed deep in the insulating regime~\cite{Simmons98}, with a
crossover to $T^{-1/3}$ behaviour near the I$\rightarrow$II
transition~\cite{SimmonsBAPS}.

Increasing the carrier density causes a transition to metallic
behaviour, in which the resistivity decreases as the temperature is
lowered. For temperatures above 0.3~K there is a weak decrease in
resistivity with decreasing temperature, although it was not possible
to ascertain the functional form of $\rho(T)$ in our earlier
data~\cite{Simmons98}. By extending the measurements to lower $T$, as
shown in Fig.~\ref{Tdeps}(b), we are able to resolve an exponential
decrease of the resistivity as $T\rightarrow 0$. This can be
quantified by fitting the low temperature ($T<T_0$) resistivity data
to the empirical formula~\cite{Pudalov97}: $ \rho(T)=\rho_0 + \rho_1
\exp(-T_0/T) $, yielding $\rho_0=0.06 h/e^2$, $\rho_1=0.006 h/e^2$,
and $T_0=0.4$~K. The exponential behaviour can be more clearly seen in
the inset to Fig.~\ref{Tdeps}(b) where [$\rho(T)-\rho_0$] is plotted
against $1/T$, revealing a linear dependence over more than a decade
change in resistivity. We note that the metallic phase appears weaker
in our samples than in other hole gases~\cite{Hanein98}, with a
smaller decrease in resistivity as $T\rightarrow 0$, and smaller
values of the ratio $\rho_1/\rho_0$ and the temperature $T_0$.

We now turn to the main result of this paper --- the observation of a
\emph{second} insulating state at high densities. As the carrier
density is further increased the metallic behaviour in regime II
becomes less pronounced, and a gradual transition to an insulating
state occurs. Fig.~\ref{Tdeps}(c) shows the behaviour of the
resistivity at the second transition, with
$p_s=1.4{\times}10^{11}\text{~cm}^{\text{-2}}$. At high temperatures
metallic behaviour is observed ($\partial\rho/\partial{T}>0$), but for
$T{<}200$~mK there is a change in sign of $\partial\rho/\partial{T}$
as a new insulating state emerges. The presence of both metallic and
insulating behaviour at the same carrier density in
Fig.~\ref{Tdeps}(c) accounts for the broadening of the transition
between the metallic (II) and insulating (III) regimes in
Fig.~\ref{fig1}(b). We noted earlier that there is also an apparent
movement of the II$\rightarrow$III transition to higher densities as
$E$ is increased. It can now be seen that this occurs because the
insulating state persists to higher temperatures as the carrier
density is increased. The strengthening of the insulator with
increasing density is shown in Fig.~\ref{Tdeps}(d), where the
percentage change in resistivity from the $T=1$~K value is plotted for
three different densities. At a density of
$1.9{\times}10^{11}\text{~cm}^{\text{-2}}$ the insulating behaviour is
visible to 500~mK, and by $2.3{\times}10^{11}\text{~cm}^{\text{-2}}$
it is observed up to $T=1$~K.

The possible origins of this insulating behaviour are now considered.
Increasing the carrier density increases $k_{F}l$ and reduces the
relative importance of many body interactions, since $r_s$ scales as
$\sqrt{1/p_s}$. Hence at sufficiently large densities the results of
earlier studies of weakly interacting systems should be
recovered~\cite{Uren,Gang4-Gorkov,AA}
--- the 2D system once again becomes insulating. In those earlier
studies two separate mechanisms were identified as causing insulating
behaviour: Weak localisation due to phase coherent backscattering, and
weak electron-electron interactions, both of which give rise to a
logarithmic correction to the Drude conductivity. To look for such a
correction the change in conductance, $\Delta\sigma
= \sigma(T)-\sigma(T=1~\text{K})$, is plotted against the temperature on a log
scale for three different carrier densities in the inset to
Fig.~\ref{Tdeps}(d). At the largest density, furthest from the
II$\rightarrow$III transition, $\log(T)$ behaviour can be observed
over a limited temperature range (150--600~mK). This data is very
similar to that observed in previous studies of weakly interacting
electron gases in lower mobility silicon inversion layers~\cite{Uren}.
At high temperatures there is a deviation from $\log(T)$ behaviour, as
phonon scattering and temperature dependent screening of impurity
scattering become important. There is also an apparent saturation of
$\Delta\sigma$ at low temperatures (which is not due to unintentional
heating effects), such that it is not possible to unambiguously
identify $\log(T)$ behaviour as the change in conductance is small
($\Delta\sigma<2\%$) and the temperature range limited.

Weak localisation can be discounted as the cause of the insulating
phase in regime III by the application of a small perpendicular
magnetic field, $B_{\perp}{=}0.1$~T. This is known to suppress phase
coherent backscattering, but we find that it does not destroy the
insulating state as would be expected if it were due to weak
localisation. Furthermore we observe a positive low field
magnetoresistance, in contrast to the negative magnetoresistance that
arises from weak localisation.

Weak electron-electron interactions cause a logarithmic correction to
the Drude conductivity which is enhanced by a magnetic field due to a
lifting of the spin degeneracy, and becomes larger with increasing
carrier density~\cite{AA}. This is in qualitative agreement with the
data in figure~\ref{Tdeps}(d), where the correction $\Delta\sigma$
becomes larger for higher carrier densities. Although the insulating
behaviour cannot be definitively ascribed to weak interactions, it is
significant that the range of temperatures over which $\log(T)$
behaviour is observed is largest for the highest densities (at which
$r_s\simeq5$), as we approach the limit for which the theory of
Ref.~\cite{AA} is applicable ($r_s\lesssim 1$).

The observation of insulating behaviour at high densities has
implications for the existence of a true 2D metal at $T=0$. The data
in Fig.~\ref{Tdeps}(d) demonstrates that as the carrier density is
reduced from $2.3{\times}10^{11}\text{~cm}^{\text{-2}}$ the onset of
insulating behaviour occurs at progressively lower temperatures. Thus
it is not certain that the metallic behaviour shown in
Fig.~\ref{Tdeps}(b) would not also become insulating at lower,
possibly inaccessible, temperatures. The corollary of this would be
that there is no true metallic state in 2D, only metallic-like
behaviour at finite temperatures. However if the insulating behaviour
saturates as $T\rightarrow0$ then it is possible that the metallic
state is stable, with a window of carrier densities within which a 2D
metal can exist. The data shown in Fig.~\ref{Tdeps}(d) do exhibit
signs of saturation below 100~mK, although we emphasise that the
functional form of $\Delta\sigma(p_s,T)$ is not yet established, and
that much lower measurement temperatures are needed before the
ultimate nature of the 2D metal can be established.

At this point it is appropriate to discuss why the second transition
to insulating behaviour at high densities has not been observed in
other studies. The heterostructure used in this work has a higher
degree of disorder than those used previously to examine the 2D
metallic state. The peak mobility is 30\% lower than in
Ref.~\cite{Simmons98} for the same $p_s$, with $r_s
\simeq 9$ at the I$\rightarrow$II transition compared to values of
11, 15 and 24 obtained in Refs.~\cite{Simmons98}, \cite{Krav} and
~\cite{SimmonsT129,Hanein98} respectively. The disorder has two
significant consequences: Firstly the metallic behaviour is weaker in
this study, as indicated by the large value of $\sigma$ at the
I$\rightarrow$II transition ($\sigma
\simeq 5e^2/h$), and the smaller values of the ratio $\rho_1/
\rho_0$ and the temperature $T_0$ (which is about half that obtained
in other hole gases at the same density~\cite{Hanein98}). Secondly the
conductivity at high densities ($\sigma \approx 100 e^2/h$) is lower
than in less disordered systems at the same density. Thus it is
possible to resolve the small change in conductivity ($\Delta\sigma
\approx e^2/h$, so $\delta\sigma / \sigma \approx 1\%$)
that identifies the second insulating regime, whereas in a less
disordered system $\delta\sigma / \sigma$ will be much smaller and
might pass undetected. However if the disorder is too large then
metallic behaviour is never observed~\cite{Popovic97}, leading to a
delicate balance between having enough disorder to observe the
insulator at high densities without destroying the metallic state.

\begin{figure}
\putfigxsz{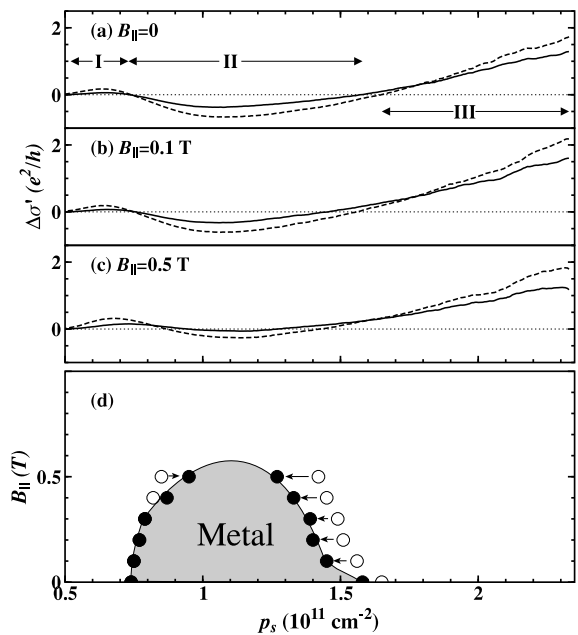}{lab}{cap}{7.5cm}
\caption{Effect of a parallel magnetic field : (a-c) Change in the differential
conductivity , $\Delta\sigma'=\sigma'(E)-\sigma'(E{=}0)$, with applied
electric field $E$, plotted as a function of carrier density for
different magnetic fields. Solid and dashed lines show data for $E=3$
and 7.5 mV/cm respectively; $T{=}30$~mK. (d)~Phase diagram for the
observation of the 2D metal as a function of carrier density and
magnetic field. The transitions are identified from the carrier
densities at which $\Delta\sigma=0$, using $E=3$ and 7.5~mV/cm for the
closed and open symbols.}
\label{Bfield}
\end{figure}

To gain further insight into the nature of the metallic and insulating
behaviour in regimes II and III we turn to the effects of a parallel
magnetic field~\cite{B_angle}. In figures~\ref{Bfield}(a-c) the change
in differential conductivity $\Delta\sigma'(E)$ due to an applied
electric field is plotted as a function of carrier density, for
magnetic fields of $B_{\parallel}{=}$0, 0.1 and 0.5~T. Once again we
divide the data into three regimes, depending on the sign of
$\Delta\sigma'$, with the points at which $\Delta\sigma'=0$ marking
the transition between insulating and metallic behaviour. Only two
values of the electric field are shown, 3 and 7.5~mV/cm, so that the
second transition (marked by vertical arrows) can be more clearly
seen. As noted previously the low density transition is independent of
$E$, whereas the second transition moves to larger densities as $E$
increases, consistent with the $T$-dependent data shown in
Figure~\ref{Tdeps}(d). It can be seen that the magnetic field
suppresses the metallic phase, reducing both the size of
$\Delta\sigma'$ and the range of carrier densities over which metallic
behaviour is observed.

We are able to construct a phase diagram for the existence of the
metal as a function of $p_s$ and $B_{\parallel}$ at $T{\simeq}30$~mK
from this data, which is plotted in Fig.~\ref{Bfield}(d). The solid
and open symbols denote the transitions identified using electric
fields of 3 and 7.5~mV/cm respectively. Initially we consider the
transitions marked by the open symbols: At low densities both $p_s$
and $\sigma$ (not shown) at the I$\rightarrow$II transition increase
as the magnetic field is applied. Conversely the II$\rightarrow$III
transition moves to lower densities and conductivities with increasing
$B_{\parallel}$, and appears to be more strongly affected by the
magnetic field than the I$\rightarrow$II transition. Thus the overall
effect of the magnetic field is to `squeeze' the metal from both
sides, until at $B_{\parallel}\gtrsim 0.7$~T insulating behaviour is
observed for all $p_s$.

Considering now the transitions identified from the smaller electric
field data ($E$=3~mV/cm, closed symbols) we see that the effect of
reducing the hole temperature is to shrink the range of densities over
which the metallic behaviour is observed. The first transition is
sharp down to the lowest hole temperatures, but the figure provides a
graphic illustration of how the metallic behaviour is extinguished
from the high density side as $E$ (or $T$) tends to zero, and it again
raises the question as to the ultimate low temperature nature of the
metallic state.

In conclusion we have observed a re-entrant
insulator--metal--insulator transition at $B=0$ in a two dimensional
system as the carrier density is increased. The re-emergence of
insulating behaviour at high carrier densities thus bridges the gap
between current data from strongly interacting systems and earlier
theoretical and experimental studies which showed that all states in a
weakly interacting 2D system are localised. Although the precise
nature of the insulating state at high densities is not fully
established, it is consistent with the interaction mechanism of
Ref.~\cite{AA} derived for weak electron-electron interactions. The
evolution of this second insulating phase with carrier density does
however raise the question as to whether the metallic behaviour at
lower densities persists to $T=0$.

We thank D.E. Khmel'nitskii, N. Cooper and C.J.B Ford for many
interesting discussions. This work was funded by EPSRC (U.K.); EHL and
DAR acknowledge support from the Isaac Newton trust and Toshiba
Cambridge Research Centre respectively.


\end{multicols}

\begin{references}
\bibitem{rsnote}We take the hole mass as $m^*=0.3m_e$;
using $0.38m_e$ will increase the values of $r_s$ calculated for this
work by 30\%.
\bibitem{B_angle}The sample was mounted at an angle of $0.35{\pm}0.05^{\circ}$
to the magnetic field.
\end{references}
\end{document}